\documentclass[noshowpacs,superscriptaddress,notitlepage,reprint]{revtex4-1}

\usepackage{epsfig,graphics,graphicx}% Include figure files
\usepackage{dcolumn}% Align table columns on decimal point
\usepackage{bm}% bold math
\usepackage{amsmath,amssymb,amsthm}
\usepackage{centernot}
\usepackage{bbold}
\usepackage{soul}
\usepackage{mathtools}
\usepackage{fullpage}
\usepackage{units}
\usepackage[usenames,dvipsnames]{xcolor}
\usepackage{braket}
\usepackage{indentfirst}
\usepackage{pgf,tikz}
\usepackage{mathrsfs}
\usetikzlibrary{arrows}
\usepackage{lipsum}
\usepackage[export]{adjustbox}

\usepackage[utf8]{inputenc}
\usepackage[T1]{fontenc}

\usepackage{dsfont}

\usepackage{mathptmx}
\usepackage{amssymb}
\usepackage{amsmath}
\usepackage{latexsym}
\usepackage{amsfonts}

\usepackage{hyperref}
\hypersetup{pdfpagemode=UseNone}

\newtheorem{theorem}{Theorem}
\newtheorem{definition}[theorem]{Definition}

\newtheorem{lemma}[theorem]{Lemma}
\newtheorem{proposition}[theorem]{Proposition}

\newcounter{rem}
\def\>{\rangle}
\def\<{\langle}

\renewcommand{\rho}{\varrho}

\def\textbf#1{{\bf #1}}

\newcommand{\pinorm}[1]{\Vert #1 \Vert_{\ell^{n}_{\infty} 
\otimes_{\pi} \ell^{n}_{\infty}}}

\def\beq{\begin{equation}}
\def\eeq{\end{equation}}
\def\beqa{\begin{eqnarray}}
\def\eeqa{\end{eqnarray}}
\def\eea{\end{array}}
\def\bea{\begin{array}}
\newcommand{\bei}{\begin{itemize}}
\newcommand{\eei}{\end{itemize}}
\newcommand{\bee}{\begin{enumerate}}
\newcommand{\eee}{\end{enumerate}}
\def\bep{\begin{proposition}}
\def\eep{\end{proposition}}
\def\bel{\begin{lemma}}
\def\eel{\end{lemma}}
\def\bet{\begin{theorem}}
\def\eet{\end{theorem}}
\def\bed{\begin{definition}}
\def\eed{\end{definition}}

\definecolor{cgreen}{RGB}{26, 199, 76}

\usepackage{soul}
\definecolor{violeta}{cmyk}{0.07,0.90,0,0.34}

\def\p{\mathbf{p}}

\def\q{\mathbf{q}}
\def\be{\begin{equation}}
\def\ee{\end{equation}}
\begin{document}
\title{Concentration phenomena in the geometry of Bell correlations}

\author{Cristhiano Duarte} 
\affiliation{International Institute of Physics, Federal University of Rio Grande do 
Norte, 59070-405 Natal, Brazil}
\affiliation{Schmid College of Science and Technology, Chapman University, One 
University Drive, Orange, CA, 92866, USA}

\author{Samuraí Brito} 
\affiliation{International Institute of Physics, Federal University of Rio Grande do 
Norte, 59070-405 Natal, Brazil}

\author{Barbara Amaral} 
\affiliation{International Institute of Physics, Federal University of Rio Grande do Norte, 59070-405 Natal, Brazil}
\affiliation{Departamento de F\'isica e Matem\'atica, CAP - Universidade Federal de S\~ao Jo\~ao del-Rei, 36.420-000, Ouro Branco, MG, Brazil}

\author{Rafael Chaves} 
\affiliation{International Institute of Physics, Federal University of Rio Grande do Norte, 59070-405 Natal, Brazil}
\affiliation{School of Science and Technology, Federal University of Rio Grande do Norte, 59078-970 Natal, Brazil}

\date{\today}

%%%%%%%%%%%%%%%%%%%%%%%%%%%%%%%%%%%%%%%%%%%%%
%%%%%%%%%%%%%Abstract%%%%%%%%%%%%%%%%%%%%%%%%
%%%%%%%%%%%%%%%%%%%%%%%%%%%%%%%%%%%%%%%%%%%%%
\begin{abstract}
Bell's theorem shows that local measurements on entangled states give rise to 
correlations incompatible with local hidden variable models. The degree of quantum 
nonlocality is not maximal though, as there are even more nonlocal theories beyond 
quantum theory still compatible with the nonsignalling principle. In spite of decades 
of research, we still have a very fragmented picture of the whole geometry of these 
different sets of correlations. Here we employ both analytical and numerical tools to 
ameliorate that. First, we identify two different classes of Bell scenarios where  the 
nonsignalling correlations can behave very differently: in one case, the correlations 
are generically quantum and nonlocal while on the other quite the opposite happens as 
the correlations are generically classical and local. Second, by randomly sampling 
over nonsignalling correlations, we compute the distribution of a nonlocality 
quantifier based on the trace distance to the local set. With that we conclude that 
the nonlocal correlations can show a concentration phenomena: their distribution is 
peaked at a distance from the local set that increases both with the number of parts 
or measurements being performed.
\end{abstract}

\maketitle

%%%%%%%%%%%%%%%%%%%%%%%%%%%%%%%%%%%%%%%%%%%%%%%%%%%
%%%%%%%%%%%%%%%%%%%%%%%%%%%%%%%%%%%%%%%%%%%%%%%%%%%
%%%%%%%%%%%%%%%%%%%%%%%%%%%%%%%%%%%%%%%%%%%%%%%%%%

%%%%%%%%%%%%%%%%%%%%%%%%%%%%%%%%%%%%%%%%%%%%%%%%%%
%%%%%%%%%%%%%%Introduction%%%%%%%%%%%%%%%%%%%%%%%%
%%%%%%%%%%%%%%%%%%%%%%%%%%%%%%%%%%%%%%%%%%%%%%%%%%

\section{Introduction}\label{sec:Intro}

Bell nonlocality \cite{Bell64} -- the fact that local measurements on some entangled 
states give rise to correlations incompatible with local hidden variable (LHV) models 
-- has become one of the cornerstones in our modern understanding of quantum theory. 
Beyond its fundamental role, it is also at the core of many relevant applications 
in information processing such as quantum cryptography 
\cite{Ekert1991,Barrett2005,Acin2006,Vazirani2014}, randomness certification 
\cite{Pironio2010,Colbeck2011}, self-testing \cite{Mayers2004,OBBDC17,CGS17}, 
dimension witnesses 
\cite{Brunner2008,KV17,CBRM16} and communication complexity problems 
\cite{Brukner2004,Buhrman2010,Chaves2012}.

Despite being more than five decades old, Bell's theorem still offers a number of 
experimental and theoretical challenges. It was only recently that the violation of a 
Bell inequality has been unambiguously confirmed experimentally 
\cite{Hensen2015,Giustina2015,Shalm2015,Rosenfeld2017}. From the theoretical 
perspective, very general frameworks have been developed 
\cite{Pitowsky1991,Brunner2014} including generalizations of Bell's original simple 
scenario -- consisting of two distant parts making two possible dichotomic measurements 
-- to more measurements, outcomes, and parts \cite{Werner2001,Collins2002,Collins2004}, 
sequential measurement scenarios \cite{Popescu1995,Gallego2014}, scenarios with 
communication 
\cite{Toner2003,Pironio2003,Maxwell2014,Chaves2015,Brask2017,Ringbauer2017,Chaves2018} 
and complex networks \cite{Chaves2016,Rosset2016}. However, still very basic questions 
remain unsolved. At the center of many open problems is the fact that there exist 
correlations agreeing with relativistic causality -- the so called 
nonsignalling (NS) correlations -- which are incompatible with quantum predictions 
\cite{Popescu1994,MBetal10}, though. Understanding how to recover the set of quantum 
correlations and more generally how to obtain a more refined picture of its relation 
to the sets of local/classical and nonsignalling correlations remains a very active 
field of research 
\cite{Van2013,Pawlowski2009,Navascues2009,Navascues2015,Fritz2013,ChavesMajenz2015}.

In a typical Bell scenario with a finite number of parts, measurements and outcomes, 
testing whether a given observed correlation falls into the nonsignalling set is 
computationally simple task, as it basically amounts to test finitely many linear 
constraints \cite{Brunner2014}. On the other hand, testing whether a correlation is 
local or quantum is considerably more difficult. It is known that the set of local 
correlations is a convex polytope \cite{Pitowsky1991}, and as such can be 
characterized by finitely many extremal points or 
equivalently \cite{LexSchrivjer99,DantzigThapa} finitely many linear inequalities (the 
non-trivial of which are known as Bell inequalities). While we can easily list such 
extremal points, obtaining the Bell inequalities is a notoriously thorny issue, the 
complexity of which grows very fast as the Bell scenario of interest becomes less 
simple. With the extremal points of the polytope we can in principle test the locality 
of a given correlation, but, once more the complexity grows very fast and indeed this 
is a problem known to be intrinsically difficult, as it stands in the NP-hard 
complexity class \cite{Pitowsky1991}. Finally, testing whether a given correlation 
admits a quantum realization is even harder. The best known solution is given by a 
hierarchy of semi-definite programs that converges asymptotically to the quantum set 
\cite{NPA07,NPA08}. The still very fragmented picture of the geometry associated with the correlations sets comes from the inherent hardness in 
characterize them \cite{Goh2018}.

One way of getting around to such difficulties while gathering at least partial 
information about the local, quantum and NS sets is to consider their relative volumes 
\cite{Cabello2005,Wolfe2012}. More recently, machine learning techniques have also been employed \cite{Canabarro2018}. Another promising venue has been to employ probabilistic 
and sampling techniques \cite{GLPV17,GGCPV16,Palazuelos15,DDO18}.
%that have acquired growing importance over the past of the years in the field of quantum information %\cite{GLPV17,GHJPV2015,MEMetal15,PSW06,BrietVidick13,BPW05,BHH16,DO12}. Not surprising they been shown to very %useful also on foundations of quantum mechanics \cite{BrietVidick13,GLPV17,DO12,DDO18,AAKDRM12} also shedding %light on the typical behavior associated with some large scale phenomena~\cite{MEMetal15,PSW06}.
%Actually, that kind of method is based on either proving that some specific object exists while without %constructing it from the scratch or based in concentration inequalities, proving that a given characteristic holds %true with high probability.
%Specifically regarding nonlocality, we have seen an increasing number of works relying on these probabilistic %approach arguments~\cite{AS04}.
For instance, in Ref.~\cite{GLPV17}, estimating both the quantum norm and the 
classical norm of random matrices with bi-orthogonally invariant probability 
distributions, the authors proved that full correlators
%\footnote{A full correlator denotes a expectation value involving observables from all parties involved in the Bell test. For instance, if one part measures $A_0$ and another part $B_0$, $\mean{A_0,B_0}$ is a full correlator but $\mean{A_0}$ is not.}
arising from local measurements on a pure bipartite quantum system are generically nonlocal.
%\ba{Eu preciso ler com mais cuidado o paper [40], mas me pareceu que eles só 
%consideram um tipo especial de
%matrizes de correlação, as chamadas \emph{bi-orthogonally invariant distributions}. Se 
%for isso mesmo, o resultado continua sendo 
%bem legal, mas a gente precisa mencionar isso dizendo que eles não exploram todo o 
%conjunto de correladores mas sim esse subconjunto
%particular. Além disso, qual medida eles estão usando? A uniforme? Também acho 
%importante deixar isso claro.}
%Employing a more refine approach~\cite{GGCPV16}, where the local dimension $d$ of each involved Hilbert space and %the number $m$ of allowed measurements per party is considered, then there exist a positive constant $\alpha_0$ %such that whenever $d/m \leq \alpha_0$ than random quantum full-correlators are nonlocal with high probability, %asymptotically in $m$, whereas if $d/m > 2$ then it is local with probability tending to 1 as $m$ goes to infinity.
If instead, the full probability distributions are considered \cite{DDO18},
%Recently, the 
%authors in~\cite{DDO18} have proven that if one focuses on behaviors rather than on correlation-matrices, and 
%assuming only a representative subset of all Bell-inequalities,
then the probability to find a $N-$partite qu$d$it system violating any Bell-inequality goes to zero, asymptotically in $N$, provided that $d>mv(2m-1)^{2}$, where $m$ is the number of measurements per party and $v$ the number of outcomes per measurement and $d$ is dimension of the associated Hilbert space.
%We refer to \cite{Palazuelos15} for a broad survey about applications of random constructions for Bell-%inequalities.  

Employing both numerical and analytical tools, in this paper we provide further 
insight into the geometry of the set of correlations arising in Bell scenarios. First, 
considering a bipartite Bell scenario with an 
increasing number $m$ of dichotomic measurements and using the probabilistic approach (as discussed, for example, in \cite{PGWPVJ08})
we prove that nonsignalling full correlators are generically nonlocal and quantum. It 
means that volume of the
local set tends to zero, whereas the volume of those correlations with quantum 
realization fills the entire nonsignalling region. Surprisingly, we also noticed that  
for a particular bipartite Bell scenario connected with the so called $n$-cycle 
scenario \cite{BC90,Klyachko2008,EntCycle,Araujo2013} quite the opposite
happens: nonsignalling correlations are generically classical.  
Second, by employing a recently introduced measure of nonlocality based on the trace distance \cite{Brito2018}, we consider a number of 
different Bell scenarios (with increasing number of outcomes, measurements, and parts) and numerically obtain the distribution of
such nonlocality quantifier over uniformly sampled NS correlations. In most cases, and perhaps not surprisingly, we obtain that
the volume of the local set decays very fast with increasing number of settings, 
outcomes or parts. However, by considering the trace distance distribution 
of such nonlocal points to the local set we have found that some interesting 
concentration-phenomena can take place.

The paper is organized as follows. Sec~\ref{sec:basics} introduces the basic 
framework and provides an overview of the results.
In Sec~\ref{sec:analytical1} and Sec~\ref{sec:analytical2} we provide analytical results regarding two scenarios where
we only take into account the full correlations produced in the experiment. 
More precisely, in Sec~\ref{sec:analytical1} we analyze a bipartite Bell scenario
with increasing number of measurements per party and in Sec~\ref{sec:analytical2} we study the geometry of correlations 
in the cycle scenario \cite{BC90,Klyachko2008,EntCycle,Araujo2013}.  
Sec.~\ref{sec:Numerics} discusses the distribution of the 
nonlocality measure introduced in \cite{Brito2018} in a variety of Bell scenarios and 
points out to the emergence of concentration phenomena.
Finally, we discuss our results as well as potential venues for future work 
in Sec.~\ref{sec:Discussion}.

%\begin{figure}
% \includegraphics[scale=0.08]{bode.jpg}
% \caption{A non-convex set. \label{fig:bode}}
%\end{figure}

%%%%%%%%%%%%%%%%%%%%%%%%%%%%%%%%%%%%%%%%%%%%%%%%%%
%%%%%%%%%%%%%%Numerics%%%%%%%%%%%%%%%%%%%%%%%%%%%%%
%%%%%%%%%%%%%%%%%%%%%%%%%%%%%%%%%%%%%%%%%%%%%%%%%%

\section{Preliminaries and an overview of the results}
\label{sec:basics}

\begin{figure}[h!]
\centering
\includegraphics[scale=1]{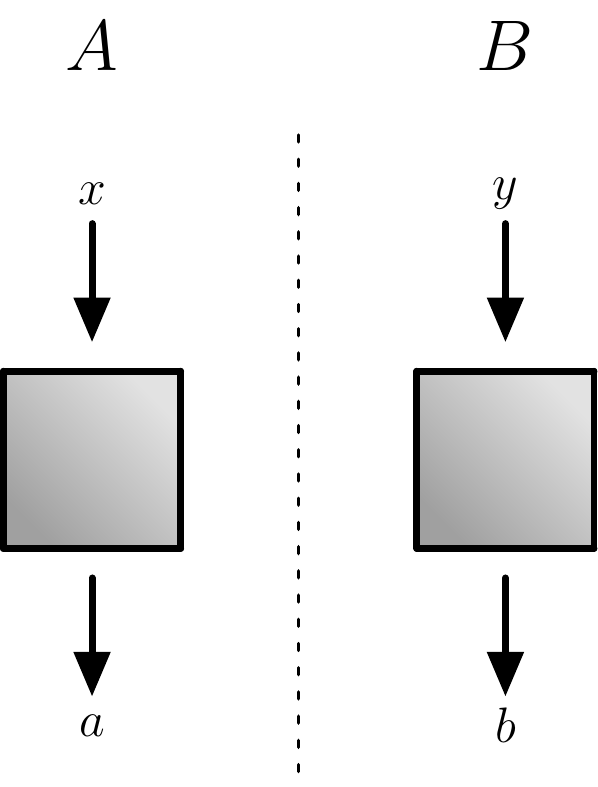}
\caption{Bipartite Bell scenario where two agents, Alice and Bob, share a pair of 
correlated measurement devices, whose inputs are labelled by $x$ and $y$ and outputs 
labelled by $a$ and $b$, respectively.}
\label{fig:bipartite}
\end{figure}

Through the paper, we will consider the paradigmatic Bell scenario 
denoted by
\begin{equation}
\Gamma:=(N,m,d), 
\end{equation}
where $N$ distant parts perform $m$ different $d$-outcome measurements on their shares 
of a joint physical system. Initially, we restrict our attention to a bipartite 
scenario (with straightforward generalization to more parts) where $N=2$ parts, Alice 
and Bob, perform measurements labeled by the variables $x$ and $y$ obtaining 
measurement outcomes described by the variables $a$ and $b$, respectively, as shown in 
Fig. \ref{fig:bipartite}. 

Assuming we do not have access to the internal mechanism of each box, our best 
description of the scenario is therefore given by the joint 
statistics $$\p \coloneqq \left\{p\left(a,b \vert x,y\right)\right\},$$ 
that is, a list containing  the probability $p\left(a,b \vert x,y\right)$ of 
obtaining outcomes $a,b$ given that the parts have measured $x,y$. In a classical 
description, based on the assumption of local realism, such correlations can be 
decomposed as
\begin{equation}
\label{LHV}
p_{\mathrm{C}}(a,b \vert x,y)= \sum_{\lambda} p(\lambda) p(a \vert x,\lambda) p(b \vert y,\lambda).
\end{equation}
This so-called \emph{local hidden variable} (LHV) description implies that all the correlations between Alice and Bob are assumed to be mediated by a common hidden variable $\lambda$ that thus suffices to
compute the probabilities of each of the outcomes, that is, $p(a\vert x,y,b,\lambda)=p(a\vert x,\lambda)$ (and similarly for $b$).

However, as discovered by Bell, there exist quantum correlations that do not comply 
with such classical description. More precisely, local measurements on entangled 
states described by a density operator $\rho$ give rise to probability distribution 
\begin{equation}
\label{quantum}
p_{\mathrm{Q}}(a,b \vert x,y)= \mathrm{Tr} \left[ \left(M^x_a \otimes M^y_b \right) \rho \right],
\end{equation}
that might violate Bell inequalities, thus precluding its explanation by LHV models.

Due to spatial distance, we do expect that outcomes of a given part are 
independent of the measurement choice of the other. This implies that the linear 
constraints
\begin{eqnarray}
\label{NS}
& & p(a \vert x) = \sum_{b} p(a,b \vert x,y)= \sum_{b} p(a,b \vert x,y^{\prime}) \\
& & p(b \vert y) = \sum_{a} p(a,b \vert x,y)= \sum_{a} p(a,b \vert x^{\prime},y),
\end{eqnarray}
known as nonsignalling (NS) conditions must hold true. Interestingly, there are NS 
correlations beyond what we can achieve with quantum mechanics 
\cite{Popescu1994,MBetal10}, that is, relativistic causality alone is not enough to 
single out the set of quantum correlations. 

Denoting the set of classical, 
quantum and nonsignalling correlations, respectively, by $\mathcal{C}_{\mathrm{C}}$, 
$\mathcal{C}_{\mathrm{Q}}$, 
$\mathcal{C}_{\mathrm{NS}}$, a fundamental result in the study of nonlocality are the strict inclusion relations:
\begin{equation}
\mathcal{C}_{\mathrm{C}} \subset \mathcal{C}_{\mathrm{Q}} \subset \mathcal{C}_{\mathrm{NS}}.
\end{equation}
Strikingly, however, is the fact that apart from these strict 
inclusions there are only few aspects known about the relation of these 
three sets of correlations. To mitigate that we employ here two different approaches 
that give us further insights about these sets of correlations.

First, we will be interested in the relative volumes between the sets. We identify two 
scenarios of interest with very different features. Considering a bipartite Bell 
scenario with an increasing number of measurement choices, we show analytically that 
generally, the set of correlations is quantum and nonlocal. That is, at the same time 
that the volume of the local set shrinks, the quantum and NS sets become arbitrarily 
close. On the opposite direction, considering the cycle scenario, again with an 
increasing number of possible measurements, we show that the volume of local 
correlations tends to unity, thus collapsing all the sets.

\begin{figure}[h!]
\centering
\includegraphics[scale=.8]{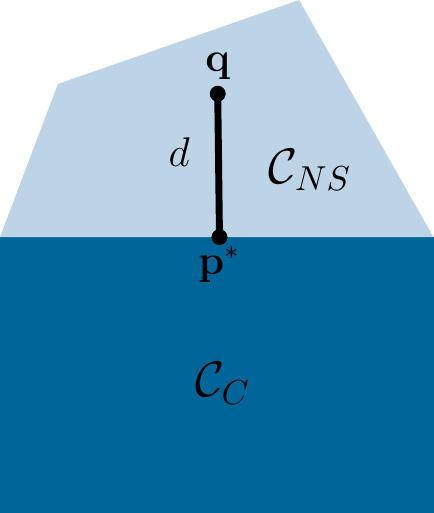}
\caption{Schematic drawing of a correlation $\q \in \mathcal{C}_{NS}$ and $d=\mathrm{NL}\left(\q\right)$, the distance (with respect to the $\ell_1$ norm) from $\q$ to the closest local correlation $\p^* \in \mathcal{C}_C$.}
\label{fig:dist}
\end{figure}

Arguably, however, the relative volume method does not reveal the full picture of what 
really happens with the set of correlations. Classifying each correlation
either as local or nonlocal is only a good indicative of what really happens 
with those sets.  By considering our second figure of merit, we show that this is 
indeed the case. We not only decided whether a given correlation is nonlocal but we 
have also  quantified its degree of nonlocality. In doing so, we employ a measure 
given by \cite{Brito2018}:
\begin{eqnarray}
\label{NLtrace}
\mathrm{NL}(\q) & & =\frac{1}{\vert x \vert \vert y \vert } \min_{\p \in \mathcal{C}_{\mathrm{C}}} \quad D(\q,\p)\\ \nonumber
& & =\frac{1}{2\vert x \vert \vert y \vert } \min_{\p \in \mathcal{C}_{\mathrm{C}}} \sum_{a,b,x,y} \vert q(a,b \vert x,y) - p(a,b\vert x,y) \vert,
\end{eqnarray}
to quantify the nonlocality of a given correlation $\q=q(a,b \vert x,y)$. This is 
the minimum trace distance between the correlation under test and the set of local 
correlations. Geometrically it should be understood (see Fig. 
\ref{fig:dist}) as how far a given dcorrelation is from the local polytope defining 
the correlations \eqref{LHV}, with the quantitative aspect that the degree 
of nonlocality of $\q$ that is equal to zero if and only if $\q$ is local. Beyond its 
geometrical and quantitative content, we point out that this distance has  
computational and numerical appeal, as it can be evaluated efficiently via a linear 
program~\cite{Brito2018}.

Considering a number of different scenarios, we have uniformly sampled over the NS 
correlations and computed the distance of each sampled correlation to the set of local 
correlations. Our numerical findings go with our analytical results described above, 
by indicating that indeed there is an unexpectedly rapid convergence. For the 
$(2,m,2)$ and $(N,2,2)$ scenarios our numerics indicates that while the relative size 
of the local set shrinks very fast (with an increasing number of measurements and 
parts, respectively), a concentration phenomenon takes part and it keeps nonlocal 
points at a distance, which increases with $m$ and $N$, from the local set. For the 
cycle  we observe a concentration of the 
volume close to the local set. Finally, for the $(2,2,d)$ scenarios with 
$d=3$ and $4$ our numerics opens two different venues for further 
investigation: either the volume of the local set increases with $d$, contrary to what 
happens with the other scenarios, or its volume decreases and shows a strong 
concentration phenomenon around it.

%%%%%%%%%%%%%%%%%%%%%%%%%%%%%%%%%%%
\section{When nonsignalling correlations can be generically quantum and nonlocal}
\label{sec:analytical1}

We explore both analytically and numerically the relative behavior between the local 
and the quantum sets of correlations with respect to the nonsignalling set when one 
considers Bell scenarios $\Gamma$ composed by two parts, $m$ measurement for each part 
and 2 possible outputs for each chosen measurement, \emph{i.e.} 
\begin{equation}
 \Gamma=(2,m,2).
\end{equation}

In addition, rather than use directly the set of correlations $\{p(a,b \vert 
x,y)\}$ as we have introduced in the previous section, it will be crucial for our findings to translate from that
probability distribution parlance to the correlators language. Therefore, assuming without loss of generality that the labels of the outcomes belong to the set $\{\pm1\}$, in this section we will only consider the following objects:
\begin{align}
  \tau_{x,y}&:=\langle x \cdot y \rangle =p(11|xy)+p(-1-1|xy) \nonumber \\ 
&-p(1-1|xy)-p(-11|xy)  \\
  \alpha_{x} &:= \langle x \rangle=p(1|x)-p(-1|x) \\
  \beta_{y}&:= \langle y \rangle=p(1|y)-p(-1|y)
  \label{Eq:Def.CorrelationMatrix}
\end{align}
for each pair of measurements $x,y \in [m]$. Although it is usual to call the matrices as in Eq.~\eqref{Eq:Def.CorrelationMatrix} as
\emph{full-correlation matrices}, we will also interchangeably use the term 
\emph{correlation matrix} to match our nomenclature with that one present in refs.~\cite{BorisTsirelson93,GLPV17} we have based
ourselves to obtain our analytical results. Actually, while these results will mostly focus on a representative subset of the
\emph{full-correlation} description of a Bell scenario (see Subsec.~\ref{SubSub:AsymptoticResults2m2}), the reader should also 
notice that our numerical findings (see Subsecs.~\ref{SubSub:NumericalResultsFullCorr2m2} and ~\ref{SubSub:NumericalResults2m2})
evidence that the same  conclusions for the local set  also hold true in the complete 
description of the $(2,m,2)$ scenario. 
\subsection{Preliminaries} \label{subsec:MathResultsPreliminaries}

Plugging the explicit expression of classical and quantum correlations in 
Eq.~(\ref{Eq:Def.CorrelationMatrix}) one can justify our next definitions of local and quantum correlation matrices:

\begin{definition}\label{Def:ClassicalCorrelationMatrix}
We say that a full-correlation matrix $\tau$ is \emph{classical} or \emph{local} whenever there is a probability 
space $(\Omega,\mathds{P})$ and response functions $A_{x},B_{y}$ with 
$A_{x}(\omega),B_{y}(\omega) \in \{-1,+1\}$ for all $\omega \in \Omega$ satisfying:
\begin{equation}
 \tau_{x,y}=\int_{\Omega}A_x(\omega)B_{y}(\omega)d\mathds{P}(\omega).
 \label{eq.DefClassicalFullCorrelations}
\end{equation}
We denote the set of all such classical matrices by $\mathcal{C}$.
\end{definition}
\begin{definition}\label{Def:QuantumCorrelationMatrix}
We say that $\tau$ is \emph{quantum} whenever there exists Hilbert spaces $H_A$ and $H_B$, a density operator $\rho$ acting on $H_A \otimes H_B$, and two families of self-adjoint operators acting on $H_A$ and $H_B$ respectively with $\max_{x,y}\left\lbrace \Vert A_x \Vert, \Vert B_y \Vert \right\rbrace \leq 1$, such that:
\begin{equation}
 \tau_{x,y}=tr[(A_x \otimes B_y) \rho].
 \label{eq.DefQuantumFullCorrelations}
\end{equation}
We denote the set of all such quantum matrices by $\mathcal{Q}$.
\end{definition}
 Roughly speaking, the next two objects we are about to describe, namely $ \Vert \cdot \Vert_{\ell^{n}_{\infty} \otimes_{\pi} \ell^{n}_{\infty}}$ the projective norm and $\gamma_{2}(\cdot)$ the $\gamma_{2}$-norm, are the most important objects within our framework. We refer to \cite{GLPV17} for all details, but the reason why we are interested in these norms in the context of correlators is due to their power in signalling out the underlying content of a given full-correlation matrix $\tau$. 
\begin{definition}
 The \emph{projective} tensor norm on $ \mathds{R}^{n} \otimes \mathds{R}^{n}$ is defined as follows:
\begin{equation}
 \Vert \tau \Vert_{\ell^{n}_{\infty} \otimes_{\pi} \ell^{n}_{\infty}} := \inf 
\left\lbrace \sum_{k=1}^{n}\Vert x_k \Vert_{\infty} \Vert y_k \Vert_{\infty} : 
\,\, \tau=\sum_{i=1}^{n}x_k \otimes y_k \right\rbrace, 
\label{eq:defprojectivenorm} 
\end{equation}
where $\tau$ is a matrix of size $n \times n$ viewed as an element of $\mathds{R}^{n} 
\otimes \mathds{R}^{n}$ 
\label{def:pi-norm}
\end{definition}
\begin{definition}
Let $\tau$ be a real $n \times n$ matrix. We define its $\gamma_2-norm$ as follows:   
\begin{equation}
\gamma_{2}(\tau):= \inf \left\lbrace \Vert X \Vert_{\ell_{2} \rightarrow 
\ell_{\infty}^{n}}\Vert Y \Vert_{\ell_{1}^{n} \rightarrow 
\ell_{2}}: \,\, \tau=XY \right\rbrace
\label{eq:defgamma2norm} 
\end{equation}
where, denoting by $R_{i}(X)$ the $i$th-row of a $n \times m$ matrix $X$ and by $C_{j}(Y)$ the $j$th-column of an $m \times n$ matrix $Y$, we have\footnote{For sake of completeness let us make clear that $\ell^{n}_{\infty}$ is the space of $n$-tuples together with $|| \cdot ||_{\infty}$ norm, and that analogously $\ell^{n}_{1}$ is also 
the space of $n$-tuples but now together with the $ || \cdot ||_{1}$ norm.}:
\begin{widetext}
\begin{equation}
 \Vert X \Vert_{\ell_{2} \rightarrow \ell_{\infty}^{n}} := \max_{i \in [n]}\Vert R_i(X)
\Vert_{2} \,\,\,\, \mbox{and} \,\,\,\, \Vert Y \Vert_{\ell_{1}^{n} \rightarrow 
\ell_{2}} := \max_{j \in [n]}\Vert C_{j}(Y) \Vert_{2}
 \label{eq:defWeirdNorms}
\end{equation}
\end{widetext}
\label{def:gamma2norm}
\end{definition}
The relation between the local and the quantum set with the norms aforementioned being given by the following set of lemmas~\cite{GLPV17}:
\begin{lemma}
 Let $\tau \in \mathds{M}_{n}(\mathds{R})$ represent a full-correlation matrix, then the following two statements hold true:
\begin{enumerate}
 \item $\tau \in \mathcal{C} \Longleftrightarrow \Vert \tau \Vert_{\ell^{n}_{\infty} 
\otimes_{\pi} \ell^{n}_{\infty}} \leq 1$
 \item $\tau \in \mathcal{Q} \Longleftrightarrow \gamma_{2}(\tau) \leq 1$.
\end{enumerate}
\label{lemma:ImportanceOfNorms}
\end{lemma}

\begin{lemma}
 Let $T$ be an $n \times n$ random matrix with bi-orthogonally invariant 
 distribution\footnote{A random matrix is said to have a bi-orthogonally invariant distribution whenever their
 probability distribution does not change when we multiply it by any orthogonal matrix, either from the right or from the left.
 Notice that it includes important sub-classes of random matrices, for instance Gaussian Matrices and those which are Haar-distributed.}, 
 and assume that 
\begin{equation}
 \exists \,\, r >0; \,\, \Vert T \Vert_{\infty} \leq \frac{(r+o(1))}{n}\Vert T 
\Vert_{1} \,\, \mbox{w.h.p. as} \,\, n\rightarrow \infty.
 \label{eq:ConditionOnSpectrumThm}
\end{equation}
Then with high probability:
\begin{equation}
 \pinorm{T} \geq \left( \sqrt{\frac{16}{15}} - o(1) \right)\gamma_2(T), \,\, \mbox{as} 
\,\, n \rightarrow \infty
\end{equation}
so that, defining 
\begin{equation}
 \tau:=\frac{T}{\gamma_{2}(T)}
\end{equation}
guarantees that it is quantum and is not classical as $n \rightarrow \infty$ with high probability.
\label{thm:MainTheoremPalazuelos}
\end{lemma}
\begin{lemma}
 Let $T$ be a random matrix with bi-orthogonally invariant distribution which fulfils Eq.~\eqref{eq:ConditionOnSpectrumThm} above. Then we have:
\begin{equation}
 \gamma_{2}(T)=(1 \pm o(1))\frac{\Vert T \Vert_{1}}{n}.
\end{equation}
\label{lemma:PalazuelosMainLemmaForQuantum}
\end{lemma}

\textbf{Remark:} On the one hand, lemma~\ref{lemma:ImportanceOfNorms} says that when we view the set of correlation matrices as a subset of $\mathds{M}_{n}(\mathds{R})$ then it is the unitary ball defined by the $\Vert \cdot \Vert_{\ell^{n}_{\infty} \otimes_{\pi} \ell^{n}_{\infty}}$ that determines whether a given correlation matrix is classical. On the other hand, the second part of the lemma guarantees that it is another unitary ball, now defined by the $\gamma_{2}$-norm, that dictates whether or not a given correlation matrix belongs to the quantum set. Both norms work therefore as a mechanism to decide about the nature of a given matrix. Despite being very technical, lemmas~\ref{thm:MainTheoremPalazuelos} and~\ref{lemma:PalazuelosMainLemmaForQuantum} 
can both be used to discuss the asymptotic behavior of correlations matrices for the $(2,m,2)$ scenario.  

\subsection{Nonsignalling correlations are generically quantum} 
\label{subsec:MathResultsMainThms}

\subsubsection{Asymptotic Results}
\label{SubSub:AsymptoticResults2m2}

Since we are interested in scenarios in which the number $n$ of inputs is greater than 1, it is possible to restate the Lemma~\ref{lemma:PalazuelosMainLemmaForQuantum} above in a weaker version, more suitable for our purposes though:

\begin{proposition}
 Let $T$ be an $n \times n$ random matrix with bi-orthogonally invariant distribution, and assume that 
\begin{equation}
 \exists \,\, r >0; \,\, \Vert T \Vert_{\infty} \leq \frac{(r+o(1))}{n^{2}}\Vert T 
\Vert_{1} 
\,\, \mbox{w.h.p. as} \,\, n\rightarrow \infty.
 \label{eq:ModifiedConditionOnSpectrumThm}
\end{equation}
Then with high probability:
\begin{equation}
 \gamma_{2}(T)=\frac{(1 \pm o(1))}{n^{2}}\Vert T \Vert_{1}.
\end{equation}
 \label{prop:ModifiedMainResultForQuantum}
\end{proposition}

Now, suppose we are focused only on those full-correlation matrices $\tau$ which are nonsignalling. In that case we have:
\begin{align}
 \Vert \tau \Vert_{\infty}  = & 1 \nonumber \\
                          \leq & (1+o(1))\frac{n^{2}}{n^{2}} \\
                            =  & (1+o(1))\frac{\Vert \tau \Vert_{1}}{n^{2}}. \nonumber
\end{align}
Therefore, in addition, accordingly to Prop.~\ref{prop:ModifiedMainResultForQuantum} we know that with high probability
\begin{align}
 \gamma_2(\tau) =& \frac{(1 \pm o(1))}{n^{2}}\Vert \tau \Vert_{1} \nonumber \\
                \leq& \frac{(1 + o(1))}{n^{2}}\Vert \tau \Vert_{1} \nonumber \\
                \leq& \frac{(1 + o(1))}{n^{2}}n^{2}  \\
                 =&   (1+o(1)).   \nonumber             
\end{align}
%
%\ba{Se a gente for mandar para um paper de física, tipo PRA, acho que temos que 
%discutir mais detalhadamente esses resultados.}
Which in turns imply that if we know beforehand that a given full-correlation matrix $\tau$ belongs to the nonsignalling set, then with high probability it also belongs to the quantum set of correlations. Summing up:
\begin{proposition}
 As the number of inputs goes to infinity, nonsignalling full-correlation matrices (with a bi-orthogonal invariant distribution) are generically quantum.
 \label{prop:NSImpliesQuantum}
\end{proposition}
Since in ref~\cite{GLPV17} the authors have found (see 
Lemma~\ref{thm:MainTheoremPalazuelos}) that quantum full-correlation matrices are generically non-classical, we can go even further. Once again, we can restate their main theorem, changing only the condition on flatness of the spectrum of $\tau$, in order to obtain:
\begin{proposition}
 Let $T$ be an $n \times n$ random matrix with bi-orthogonally invariant distribution, and assume that 
\begin{equation}
 \exists \,\, r >0; \,\, \Vert T \Vert_{\infty} \leq \frac{(r+o(1))}{n^{2}}\Vert T 
\Vert_{1} \,\, \mbox{w.h.p. as} \,\, n\rightarrow \infty.
\end{equation}
Then with high probability:
\begin{equation}
 \pinorm{T} \geq \left( \sqrt{\frac{16}{15}} - o(1) \right)\frac{\Vert T 
\Vert_{1}}{n^{2}}, \,\, \mbox{as} 
\,\, n \rightarrow \infty
\end{equation}
\end{proposition}
With that result in hands, assuming that the $\tau$ is a random full-correlation matrix belonging to the nonsignalling set one has:
\begin{equation}
  \pinorm{T} \geq \left( \sqrt{\frac{16}{15}} - o(1) \right), \,\, \mbox{as} 
\,\, n \rightarrow \infty,
\end{equation}
which in turns shows that generically nonsignalling correlations do not belong to the set of classical correlations. Summing up:

\begin{proposition}
 As the number of inputs goes to infinity, nonsignalling full-correlation matrices (with a bi-orthogonal invariant distribution) are generically non-classical.
\end{proposition}

Giving to the reader a less abstract glimpse of our findings, we finish the present section discussing numerically what happens with the distance distribution for nonlocal correlations in many different $(2,m,2)$ scenarios.    

%%%

\subsubsection{Numerical results for the $(2,m,2)$ Bell scenario }
\label{SubSub:NumericalResults2m2}

% The 
% nonsignalling set is a 
% $n^2+2n$-dimensional set in $\mathds{R}^{4n^2}$. In order to perform the sampling in 
% $\mathcal{C}_{NS}$, we need to find a full-dimensional parametrization of 
% $\mathcal{C}_{NS}$ 
% in $\mathds{R}^{n^2+2n}$. With this purpose, we write the behaviours 
% in terms of expectation values instead of probability distributions. We define
% \begin{eqnarray}
% \left\langle x \right\rangle&=&p(1\vert x) - p(-1\vert x)\\
% \left\langle y \right\rangle&=&p(1\vert y) - p(-1\vert y)\\
% \left\langle xy\right\rangle &=&p(11\vert xy) + p(-1-1\vert xy) \nonumber\\
% & & - p(-11\vert xy) - p(1-1\vert xy).
% \end{eqnarray}
% The map that takes each distribution $p\left(ab\vert xy\right)$ in the vector 
% \be\left(\left\langle x \right\rangle, \left\langle y \right\rangle,\left\langle xy\right\rangle\right)\ee
%  is bijective and gives a full-dimensional parametrization of the nonsignalling set 
% in $\mathds{R}^{n^2+2n}$.
% Notice that the image of the nonsignalling set under this map belongs to the 
% hypercube $\left[-1,1\right]^{n^2+2n}$. A point in 
% the hypercube belongs to the image of the nonsignalling set if it satisfies
% \begin{equation}
%  p(a,b|x,y)=\frac{ 1+a\langle A_x \rangle + b\langle B_y \rangle + ab\langle 
% A_xB_y \rangle}{4} \geq 0
% \end{equation}
% for every $a,b = \pm 1$.

Moving on, now we analyse what happens when we estimate the relative volume of the 
classical set when considering the complete correlation
\be\left(\left\langle x \right\rangle, \left\langle y \right\rangle,\left\langle  
xy\right\rangle\right).\ee For doing so we $i)$ used the Gibbs sampler within the  
MATLAB  function \emph{cprnd} as implemented by Benham \cite{cprnd} to $ii)$ generate 
a set of uniformly distributed points in $\q \in \mathcal{C}_{NS}$ and $iii)$ then 
compute $\mathrm{NL}(\q)$.
Table \ref{table:2n2} summarizes our results. The case $m=2$ corresponds to the usual 
Clauser-Horne-Shimony-Holt scenario, and what we have found coincide with the  
results of Refs.~\cite{Cabello2005,Wolfe2012}, though there the authors have based 
themselves upon other methods.
The volume of the classical set decreases  fast with $m$, as we can see in the 
Fig.~\ref{fig:2m2_vol}a. Notice, however, that the decaying in the volume ratio is 
much more notable in the scenario where only full correlators are considered, thus 
showing the relevance of marginal information in the geometry of the Bell 
correlations. 

\begin{table}[h!]
\centering
\begin{tabular}{|c|c|c|c|c|}
\hline
m & $10^x$ & $\sharp$L    & \%L            \\ \hline
2 & 7 & 9414201 & 94.14  \\ \hline 
3 & 6 & 621123  & 62.11        \\ \hline
4 & 6 & 212093  & 21.20        \\ \hline
5 & 6 & 37396    & 3.73 \\ \hline
\end{tabular}
\caption{Volume of the classical set in the $(2,m,2)$ scenario  with the Gibbs sampler within the MATLAB function crpnd for $m=2,3,4,5$. The second column shows the number of points sampled, $\sharp$L denotes the number of local points found (within numerical precision of $10^{-10}$) and $\%$L denotes the ration between the number of local points and the number of nonsignalling points.}
\label{table:2n2}
\end{table}
%5 & 5 & 7226    & 7.23    \\ \hline

As in the case with full correlators only (see Subsec. \ref{SubSub:NumericalResultsFullCorr2m2}), the probability of finding a point at 
distance $l$ from the local set does not decrease monotonically with $l$, except for 
$m=2$. For $m=2$, each nonlocal extremal point violates only one facet-defining Bell 
inequality and hence the nonlocal portion of the nonsignalling set consists of disjoint 
pyramids whose apex is one of the nonlocal extremal points and the basis is a simplex 
whose vertices are the local extremal points saturating the corresponding inequality. 
This is no longer the case for $m>2$, where there are correlations that might violate 
more than one facet-defining inequality. The nonsignalling and classical sets have a 
much more complicated geometry in this case, which manifests as the non monotonic 
behavior of the trace distance shown in Fig. \ref{fig:2m2}a.

\subsubsection{Numerical results for the full-correlation $(2,m,2)$ Bell scenario }
\label{SubSub:NumericalResultsFullCorr2m2}

In the full-correlation framework, the nonsignalling  set for the $(2,m,2)$ scenario is the hypercube $\left\{-1,1\right\}^{m^2}$. 
To estimate the relative volume of the classical set we sample uniformly from 
this hypercube and calculate the trace distance $\mathrm{NL}(\q)$. We summarize our 
findings in Table \ref{table:2m2full}. These results show that the 
volume of the local set decreases rapidly as $m$ grows, supporting the analytical 
results obtained in the previous section.

\begin{table}[h!]
\centering
\begin{tabular}{|l|l|l|}
\hline
$m$  & $\sharp L$     & $\%L$          \\ \hline
2  & 666657 & 66.66    \\ \hline
3  & 140138 &  14.01 \\ \hline
4  & 8470  &   0.847 \\ \hline
5  & 165  & 0.016             \\ \hline
\end{tabular}
\caption{Volume of the classical set in the full-correlation  $\left(2,m,2\right)$ scenario. 
From a total of $10^6$ sampled points,  $\sharp L$ denotes the number of local points found (within numerical precision of $10^{-10}$)
and $\% L$ denotes the ratio between the number of local points and the number of  nonsignalling points.}
\label{table:2m2full}
\end{table}

\begin{figure}
 \centering
 \includegraphics[scale=0.5]{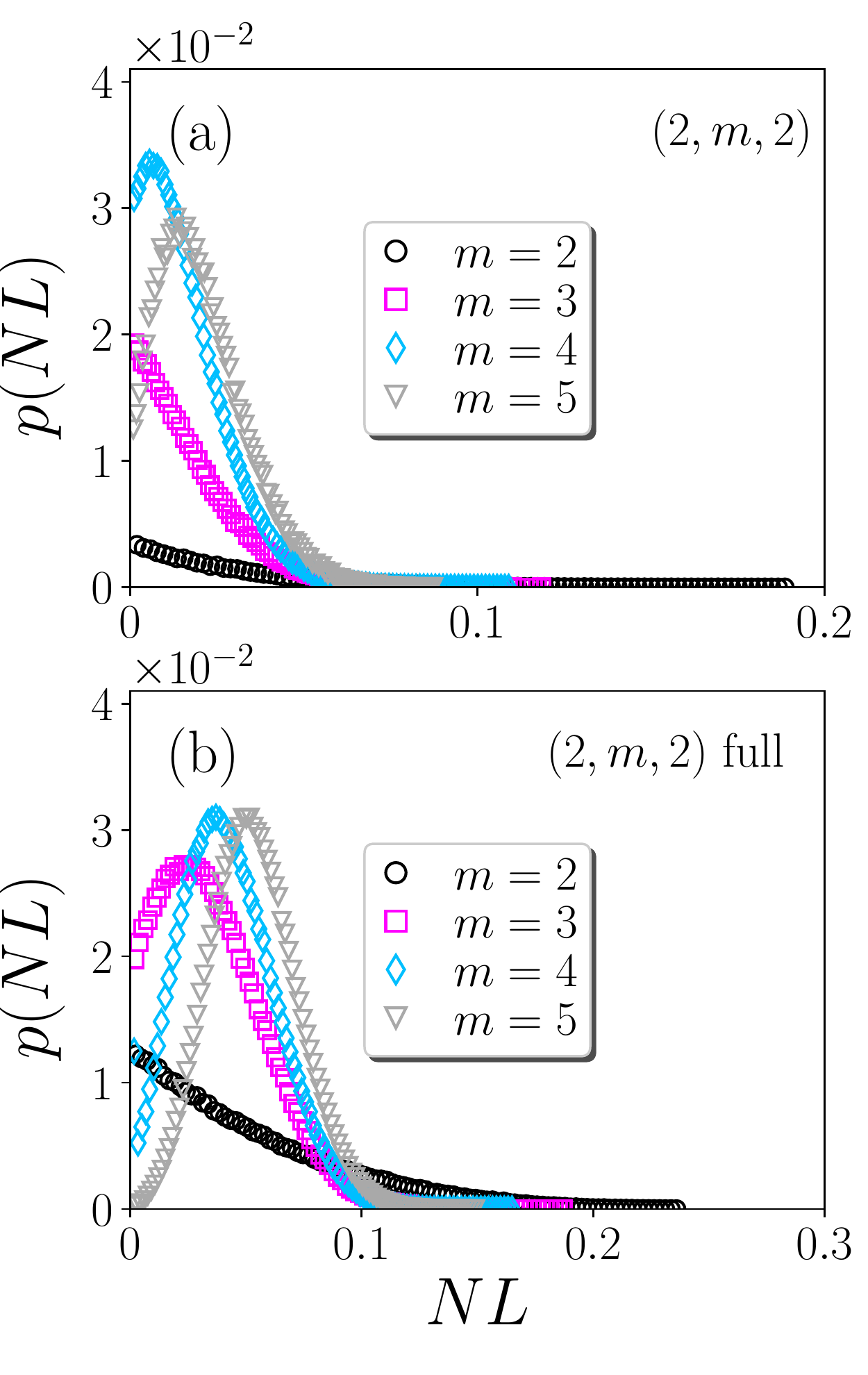}
 \caption{Distance distribution for the nonlocal correlations in the Bell scenario with $10^6$ samples. $a)$ The $(2,m,2)$ Bell scenario for $m=2,3,4,5$. $b)$ $(2,m,2)$ Bell scenario where only full correlators are considered for $m=2,3,4,5$. The numerical precision is within of $10^{-10}$. We are considering local if the result of $\mathrm{NL}(\q) \leq 10^{-10}$ and nonlocal otherwise.}
 %\label{fig:2m2full}
 \label{fig:2m2}
\end{figure}

\begin{figure}
 \centering
 \includegraphics[scale=0.45]{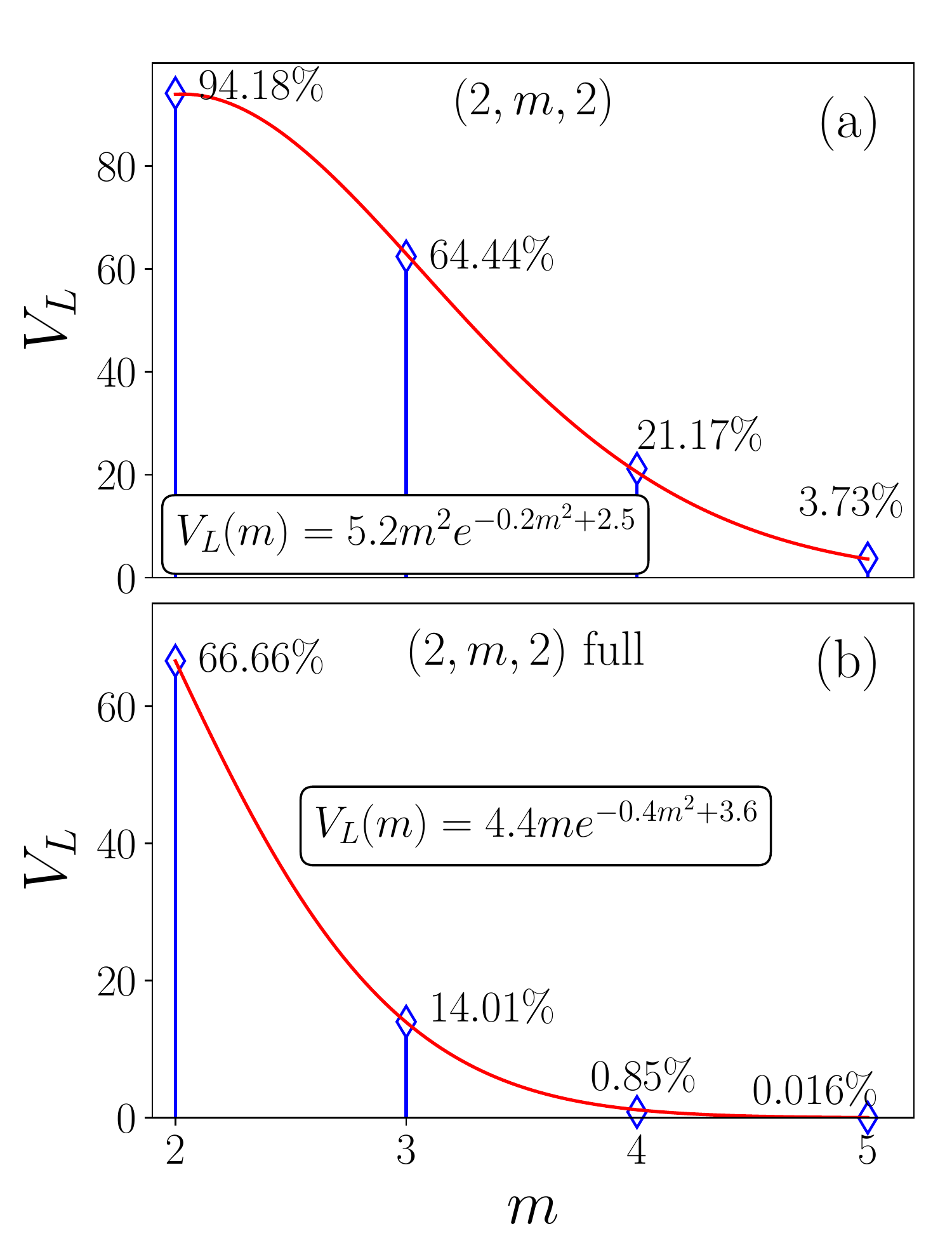}
 \caption{ $a)$ Relative volume $V_\mathrm{L}/V_{\mathrm{NL}}$ for the $(2,m,2)$ Bell. $b)$ Relative volume 
$V_\mathrm{L}/V_{\mathrm{NL}}$ for the $(2,m,2)$ Bell scenario with only full correlators. In both cases the 
relative volume of the local set decreases with $m$, however for the full correlators scenario the volume decreases faster. The red line is the best fitting we numerically find for this points.}
  \label{fig:2m2_vol}
\end{figure}

The distance distribution for the nonlocal full correlators for $m=2,3,4,5$ is shown in 
the Fig.~\ref{fig:2m2}b, and in Fig.~\ref{fig:2m2_vol}b we can see how the relative 
volume of the local set decreases with $m$. 
The probability of finding a point at distance $l$ from the local set decreases monotonically with $l$ only for $m=2$. 
For $m>2$ we have the emergence of a concentration- like phenomenon, where the distribution shows a peak 
that moves to the right with increasing $m$. That is, not only the volume of the local set decreases fast, most of the nonlocal 
points concentrate a certain distance that increase with the number of measurement settings $m$. Clearly, we have here a signature 
of the complicated geometry of Bell correlations.

%\begin{figure}
 %\centering
 %\includegraphics[scale=0.4]{2m2_nonlocal_histogram.eps}
 %\caption{Distance distribution for the nonlocal behaviours in the full correlation $\left(2,m,2\right)$ Bell scenario for $m=2,3,4,5$ and  $10^6$ samples. \sam{The numerical precision of the results are given with $16$ decimal places. We are considering local if the result of $\mathrm{NL}(\q) \leq 10^{-10}$ and nonlocal otherwise.}}
  %\label{fig:2m2full}
%\end{figure}

%%%

\section{Nonsignalling correlations can be generically local}
\label{sec:analytical2}

Here we will give an example of a scenario where the relative volume of the set of local correlations tends to unity. The scenario in question is known as the cycle scenario: we have $n$ measurements $x_1, \ldots ,  x_n$ for the first party and $n$ measurements $y_1, \ldots ,  y_n$ for the second party, with binary outputs $\pm 1$. We consider only the joint distributions for $x_i$ and $y_j$ whenever $i-j=1 \mod n$.

Similarly to a usual Bell scenario, the cycle scenario is described by the probability distributions
\be
p\left(a_ib_j \vert x_iy_j \right)
\ee
for $i-j=1 \mod n$. 

In what follows we will analytically compute the volume of local  and nonsignalling 
sets as a function of $n$ and show that $V_{\mathrm{L}}/V_{\mathrm{NS}} \rightarrow 1$. 
It is crucial for our construction to consider exclusively the full correlators 
$\left\langle x_i y_j \right\rangle$. However, at the end of the section, we will 
provide  evidence indicating that the same result holds also considering the 
complete probability distributions.

\begin{figure}
 \centering
 \includegraphics[scale=0.4]{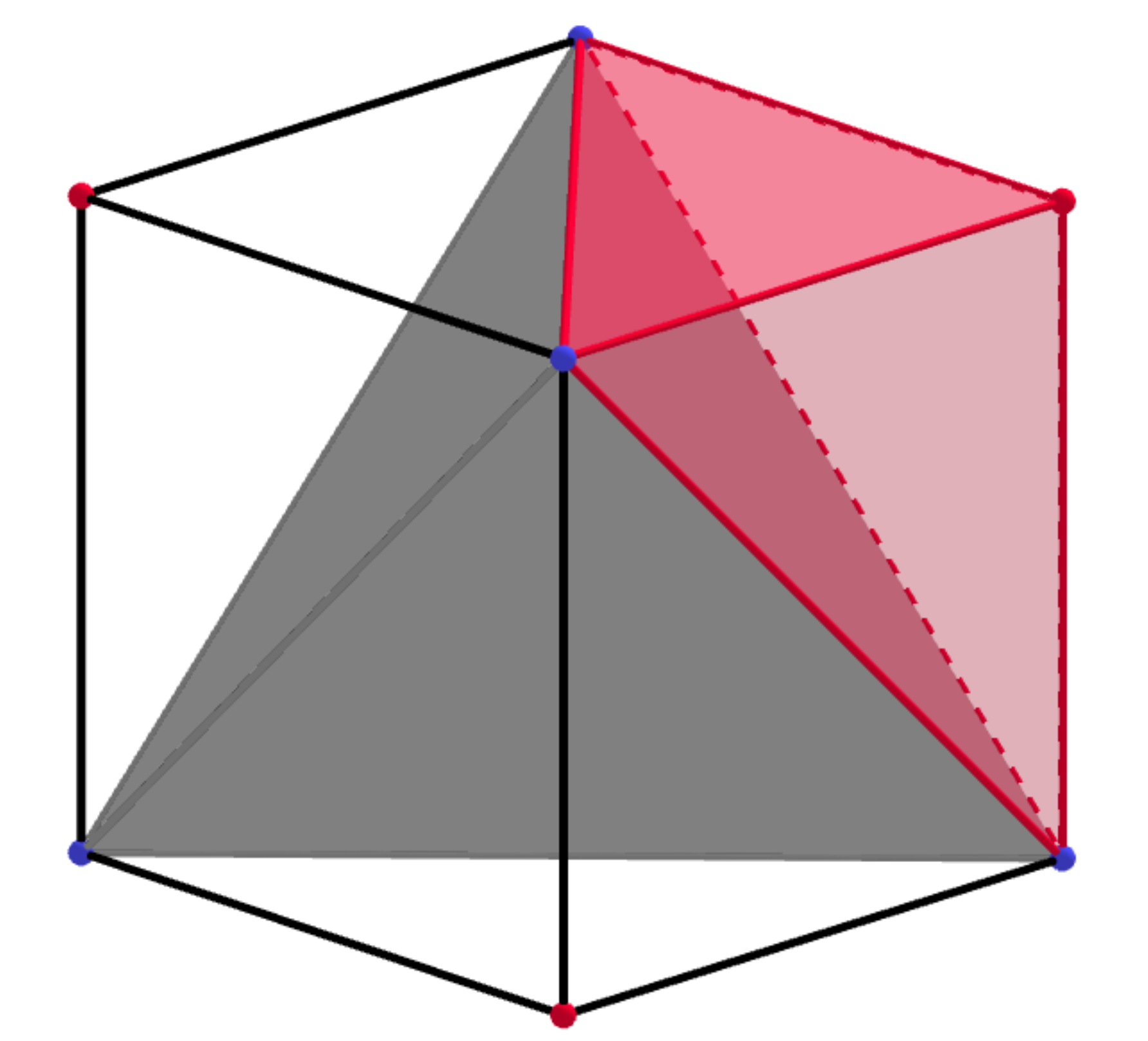}
  \caption{Schematic representation of the set of nonsignalling correlations in the 
cycle scenario. The vertices in blue represent the local vertices, whose convex rule 
gives the local set, depicted in gray. The vertices in red represent the nonlocal 
vertices. Notice that the nonlocal part of the nonsignalling set corresponds to 
disjoint pyramids, one of which is shown in red.}
  \label{fig:c3full}
\end{figure}

In the full-correlation framework, the nonsignalling set is  the hypercube $\left\{-1,1\right\}^{2n}$ with $2^{2n}$ vertices, $2^{2n-2}$ of them nonlocal (odd number of entries equal to $-1$) and the other $2^{2n-1}$ corresponding to local ones (even number of entries equal to $-1$) \cite{Araujo2013}. 
Each nonlocal extremal point violates only one facet-defining Bell inequality and hence the nonlocal portion of the nonsignalling set consists of $2^{2n-1}$ disjoint pyramids, whose apex is one of the nonlocal extremal points and the basis is a simplex in dimension $2n-1$ whose vertices are the $2n$ local extremal points saturating the corresponding inequality. Thus, the volume of the local set is the volume of the hypercube minus the volume of these pyramids. A schematic representation of this sets is shown in Fig. \ref{fig:c3full}.

Remembering basic geometry, we know that to calculate the volume of one of these pyramids we need to know the area of the basis and the height. First, we calculate the area of the basis. The volume of a simplex with $v+1$ vertices and side $s$ is equal to
\begin{equation}
V_S= \frac{s^v}{v!}\sqrt{\frac{v+1}{2^v}}.
\end{equation}
The basis of the pyramid is a simplex with $2n$ vertices and side $\sqrt{8}$,  the Euclidean distance  from two adjacent vertices of the local polytope. Hence, its area is given by
\be
A_B= \frac{\sqrt{8}^{2n-1}}{(2n-1)!}\sqrt{\frac{2n}{2^{2n-1}}}.
\ee
In turn, the height of the pyramid is the Euclidean distance of the nonsignalling extremal point to the uniform 
convex mixture of the local ones, and it is given by
\be
h=\frac{2}{\sqrt{2n}}.
\ee
We can now calculate the volume of the pyramid simply as
\be
V_P=\frac{h\times A_B}{2n}=\frac{2^{2n}}{(2n)!}.
\ee
Since we have $2^{2n-1}$ such pyramids, the total volume will be
\be
V_{L}=2^{2n}-2^{2n-1} \times \frac{2^{2n}}{(2n)!} = 2^{2n}\left(1-\frac{2^{2n-1}}{(2n)!}\right).
\ee
Clearly, with $n \rightarrow \infty$ the volume of the local set tends to the volume of the hypercube and thus $V_{\mathrm{L}}/V_{\mathrm{NS}} \rightarrow 1$. As can be seen from Table~\ref{table:n_cycle}, the relative volume of the local set tends very rapidly to unity and already at $n=4$ it achieves $\approx 0.997$.

%We have also computed the relative volume sampling from the hypercube and using the trace distance. The results are shown in 
%table \ref{table:n_cycle_full}.

\begin{figure}
 \centering
 \includegraphics[scale=0.5]{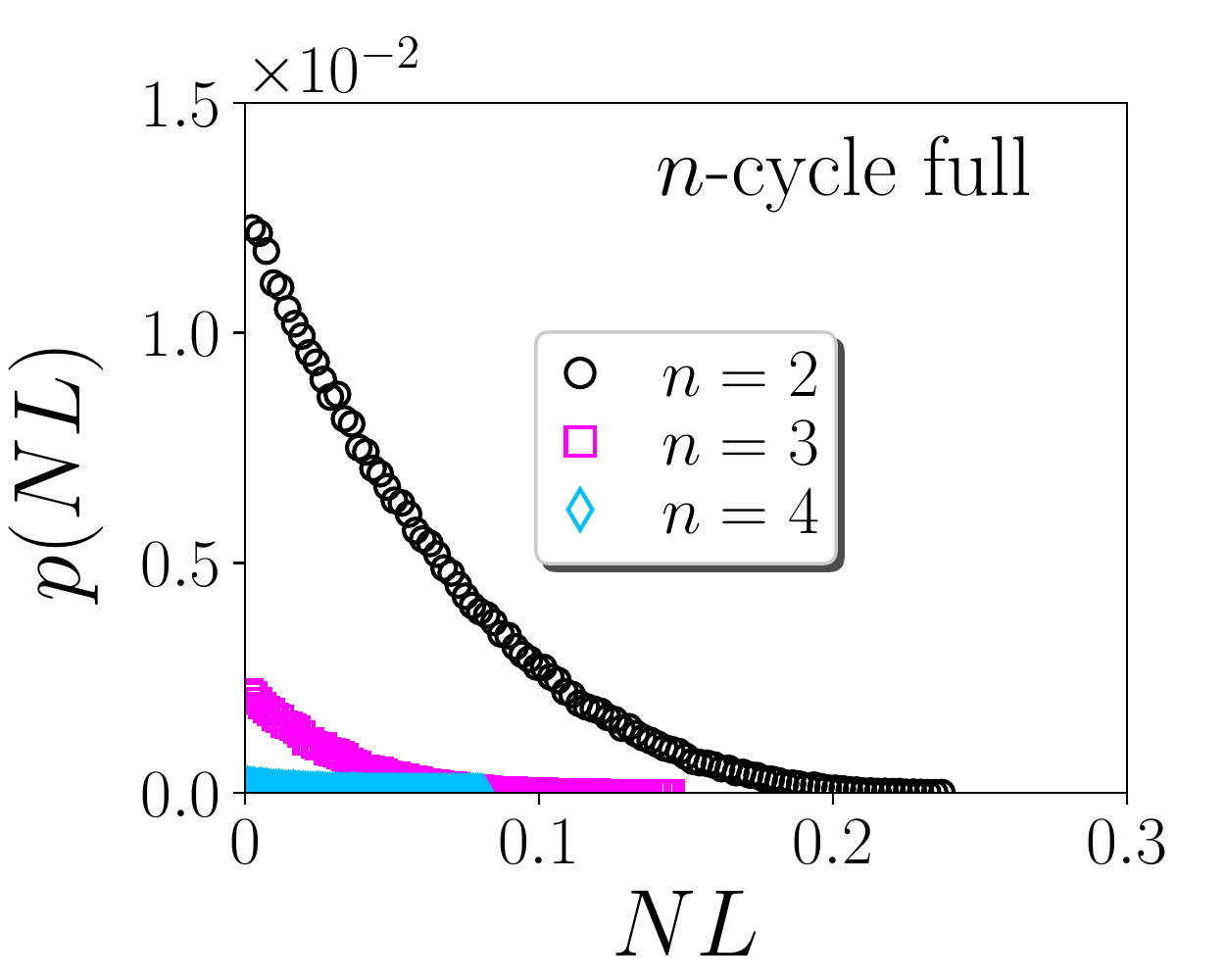}
 \caption{Distance distribution for the  contextual full correlators in the $n$-cycle scenario for $n=2,3,4$ and $10^6$ samples.
% Distance distribution for the  contextual correlators in the $n$-cycle scenario for $n=2,3$ and $10^6$ samples.
 The numerical precision is within of $10^{-10}$. We are considering local if the result of $\mathrm{NL}(\q) \leq 10^{-10}$ and nonlocal otherwise.}
  \label{fig:n-cycle}
\end{figure}

\begin{widetext}

\begin{table}[]
\begin{center}
\begin{tabular}{|l||l|l||l|l|l|l|l||l|l|}
\hline
& \multicolumn{2}{c||}{Full} & \multicolumn{5}{c||}{Complete M1}& \multicolumn{2}{c|}{Complete M2}\\ \hline
n    & $\sharp L$ & $\%L $ &   $\sharp P=10^x$  &$\sharp NS$& $\sharp L$ &$\% NS$&  $\%L $   & $\sharp L$ & $\%L $ \\ \hline
%3 & 334131 & 0,334  & 0,333  \\ \hline
2 & 666362   & 66,7\% & 7 &268896&253061&2,69&94,11 & 939828&93,98\\ \hline
%5 & 866691 & 0,867  & 0,867  \\ \hline
3 & 955538  & 95,6\% & 8 &442814&442589&0,44&99,95 & 999617&99,96\\ \hline
%7 & 987370 & 0,987  & 0,987  \\ \hline
4 & 996816  & 99,7\% & 9 & 727331&727328&0,07&100 & 999997 &$\sim$ 100\\ \hline
%9 & 999319 & 0,999  & 0,999  \\ \hline
\end{tabular}
\end{center}
\caption{Volume of the set of local correlators in the cycle scenario for  
$n=2,3,4$. The first column shows the value of $n$.
Columns 2 and 3 show the number of full-correlation local points $\sharp L$ and the percentage of full-correlation local points
$\% L$, respectively, from a sample of $10^6$ points for every $n$. The data obtained when we sample considering the entire correlation with methods M1 (sampling over the hypercube and 
discarding the signalling points) and M2 (sampling directly over the NS polytope using the MATLAB cprnd function)
is show in columns 4-10. Columns 4-8 
show the number of points $\sharp P$ sampled, the number of nonsignalling points $\sharp NS$, the number of local 
points $\sharp L$, the percentage of nonsignalling points $\%NS$ and the percentage of local points $\%L$ obtained with method M1.
Columns 9 and 10 show the number of  local points $\sharp L$ and the percentage of  local points
$\% L$, respectively, from a sample of $10^6$ points, obtained with method M2. }
\label{table:n_cycle}
\end{table}

\end{widetext}
%\begin{figure}
% \centering
% \includegraphics[scale=0.6]{fig_n_cycle_full.eps}
% \caption{Volume of the noncontextual full correlators in the $n$-cycle scenario as a function of $n$.}
% \label{fig:n_cycle}
%\end{figure}
%
%
To complete the picture of the cycle with full correlators, we have uniformly sampled inside the nonsignalling set and computed the distance to the set of local correlations. The results are shown in Fig.~\ref{fig:n-cycle}.

%We can quantify the contextuality of a behavior $\mathrm{B}$ using the trace distance
%\begin{eqnarray}
%\label{Ctrace}
%\mathrm{C}\left(\mathrm{B}\right) & & =\frac{1}{n } \min_{\mathrm{B}^*\in \mathcal{NC}} \quad \mathrm{D}\left(\mathrm{B},\mathrm{B}^*\right)\\ \nonumber
%& & =\frac{1}{n } \min_{\mathrm{B}^* \in \mathcal{NC}} \sum_{a_i,a_{j}, x_i, x_{j}} \vert p(a_i,a_j \vert x_i,x_j) - p(a_i,a_j\vert x_i,x_j) \vert. 
%\end{eqnarray}
%where $\mathcal{NC}$ is the set of noncontextual behaviours and  $i-j=1 \mod n$.
%We have that $\mathrm{C}\left(\mathrm{B}\right)=0$ iff $\mathrm{B}$ is noncontextual and the value of 
%$\mathrm{C}\left(\mathrm{B}\right)$ is a quantitative measure of the degree of contextuality of $\mathrm{B}$.

%\begin{figure}
 %\centering
 %\includegraphics[scale=0.38]{ncyclefull.eps}
% \includegraphics[scale=0.6]{graph_ncycle_full.eps}
% \caption{(colours on-line) Distance distribution for the  contextual full correlators in the $n$-cycle scenario for $n=2,3,4$. \sam{grafico ta muito vazio nao acham?}}
%  \label{fig:graph_n_cycle_full}
%\end{figure}

Finally, we have also computed the distribution of the trace distance taking into account an uniform sampling over the complete probability distribution. We use two different methods to sample the points from the nonsignalling set. The first step in both methods is to write the probability distribution in terms of expectation values:
\begin{eqnarray}
\left\langle x_i \right\rangle&=&p(1\vert x_i) - p(-1\vert x_i)\\
\left\langle y_j \right\rangle&=&p(1\vert y_j) - p(-1\vert y_j)\\
\left\langle x_iy_j\right\rangle &=&p(11\vert x_iy_j) + p(-1-1\vert x_iy_j) \nonumber\\
& & - p(-11\vert x_iy_j) - p(1-1\vert x_iy_j).
\end{eqnarray}
The map that takes each nonsignalling probability in the vector
\be
\left(\left\langle x_i \right\rangle, \left\langle y_j \right\rangle, \left\langle x_iy_j\right\rangle\right)
\ee
with $i-j=1 \mod n$ is bijective and gives a full-dimensional parametrization of the nonsignalling set in $\mathds{R}^{4n}$. Further, notice that the image of the nonsignalling set under this map belongs to the hypercube $\left[-1,1\right]^{4n}$. A point in the hypercube belongs to the image of the non-disturbing set if it satisfies \be p\left(a_ib_j\vert x_iy_j\right)=\frac{ 1+a_i\langle x_i \rangle + b_j\langle y_j \rangle + a_ib_j\langle x_iy_j\rangle}{4} \geq 0 \label{eq:condND} \ee for all $a,b =\pm 1$ and $i-j=1 \mod n$.
In the first method, we sample uniformly from the hypercube $\left[-1,1\right]^{4n}$, 
test if inequalities \eqref{eq:condND} are satisfied, and discard the points that are 
outside the image of the nonsignalling set. Then we compute the distance of this point 
to the local set. The results are shown in Table \ref{table:n_cycle}. 
Notice that this method is very inefficient because the relative volume of the 
nonsignalling set with respect to the hypercube is very small. Nevertheless
it offers a good testing ground for other methods since we are sure that the sampling is indeed uniform.

% \begin{table}[h!]
% \begin{center}
% \begin{tabular}{|c|c|c|c|c|c|}
% \hline
% n & $\sharp P=10^x$ & $\sharp NS$      & $\sharp L$       & $\%NS $& $\%L$           \\ 
% \hline
% %3 & 8                      & 6669733 & 4445530 & 6,67 & 66,65      \\ \hline 
% 2 & 7                      & 268896  & 253061  & 2,69 & 94,11       \\ \hline
% %5 & 8                      & 1092816 & 1085711 & 1,09 & 99,35      \\ \hline 
% 3 & 8                      & 442814  & 442589  & 0,44 & 99,95      \\ \hline
% %7 & 8                      & 179654  & 179652  & 0,18 & 100,00      \\ \hline
% 4 & 9                      & 727331  & 727328  & 0,07 & 100,00     \\ \hline
% \end{tabular}
% \end{center}
% \caption{Volume of the noncontextual set in the cycle scenario sampling from the hypercube. In that table $\sharp P$ is the number of points sampled from the hypercube, $\sharp NS$ is the number of nonsignalling points, $\sharp L$ is the number of local points, $\%NS$ is the relative frequency of  nonsignalling points and $\%L$ is the relative frequency of local points. }
% \label{table:n_cycle}
% \end{table}
% 
In the second method we use the Gibbs sampler within the  MATLAB  function \emph{cprnd} \cite{cprnd} .
The results are shown in Table \ref{table:n_cycle}
and agree with the results obtained with the previous method. The volume of the local set grows with $n$, showing the same behavior as in
 Fig. \ref{fig:n-cycle}.
% 
% \begin{table}[h!]
% \begin{center}
% \begin{tabular}{|c|c|c|c|c|}
% \hline
% $n$ & $\sharp \mathrm{NS}=10^x$ & $\sharp \mathrm{L}$      & $\%\mathrm{L}$    
%   \\ \hline
% %3 & 6                       & 667231 & 66,72  \\ \hline
% 2 & 6                       & 939828 & 93,98   \\ \hline
% %5 & 6                       & 993351 & 99,34   \\ \hline
% 3 & 6                       & 999617 & 99,96   \\ \hline
% %7 & 6                       & 999975 & 99,99         \\ \hline 
% \end{tabular}
% \end{center}
% \caption{Volume of the local set in the $n$-cycle scenario  with the Gibbs sampler within the MATLAB function crpnd.}
% \label{table:n_cycle_2}
% \end{table}

The probability of finding a point at distance $D$ from the local 
set decreases monotonically with $D$. This is a signature of the geometry of the 
nonsignalling set. Again, in this scenario, each nonlocal extremal point violates only 
one facet-defining Bell inequality and hence the nonlocal portion of the nonsignalling 
set consists of disjoint pyramids whose apex is one of the nonlocal extremal points and 
the basis is a simplex whose vertices are the local extremal points saturating the 
corresponding inequality.

%%%%%%
\section{Concentration phenomena of nonlocal correlations}
\label{sec:Numerics}
So far we have focused on a bipartite scenarios where the number of measurements 
increase but the number of outcomes is always two. In the following we show that while 
similar results hold true when we increase the number of parts, keeping 
dichotomic measurements, they can change dramatically if instead we keep fixed the 
number of parts and measurements and increase the number of measurement 
outcomes.

%%%

\subsubsection{$\left(N,2,2\right)$ Bell Scenario}
%\sam{B\'arbara, falta a figura desse cen\'ario? N\~ao tenho os dados para fazer.}
In the $\left(N,2,2\right)$ Bell scenario $N$ parts have each two binary measurements. 
To find a full-dimensional parametrization of $\mathcal{C}_{NS}$ we write the correlations in terms of the probabilities 
\begin{align}
p\left(-1\vert x_i\right), \nonumber \\
p\left(-1-1\vert x_ix_j\right), \nonumber\\
p\left(-1-1-1\vert x_ix_jx_k\right),\nonumber \\
p\left(-1-1\ldots -1\vert x_1x_2\ldots x_N\right),
\label{eq:N22}
\end{align}
We can recover every probability $p\left(a_1\ldots a_N\vert x_1 \ldots x_N\right)$ from the  vector with entries given by equations \eqref{eq:N22}.
Hence, the map that takes each correlation to the corresponding vector \eqref{eq:N22} 
is bijective and gives a full-dimensional parametrization of the nonsignalling set.
Notice that the image of the nonsignalling set under this map belongs to the hypercube 
with coordinates in $[0,1]$. A point in the hypercube belongs to the image of the 
nonsignalling set if every $p\left(a_1\ldots a_N\vert x_1 \ldots x_N\right)$ recovered 
from it is positive.

To estimate the relative volume of the local set we use the 
Gibbs sampler within the \emph{crnpd} MATLAB function and calculate trace distance $\mathrm{NL}$.
The results are shown in Table \ref{table:N22}.
\begin{table}[h!]
\centering
\begin{tabular}{|l|l|l|l|}
\hline
$N$ & $\sharp NS=10^x$ & $\sharp L$     & $\%L$          \\ \hline
2 & 7 & 9414201 & 94.14201  \\ \hline
3 & 6 & 585206 & 58.52  \\ \hline
4 & 6 &   40576  &  4.06  \\ \hline
\end{tabular}
\caption{Volume of the local set in the  correlation  $\left(N,2,2\right)$ scenario. }
\label{table:N22}
\end{table}

\begin{figure}
 \centering
 \includegraphics[scale=0.45]{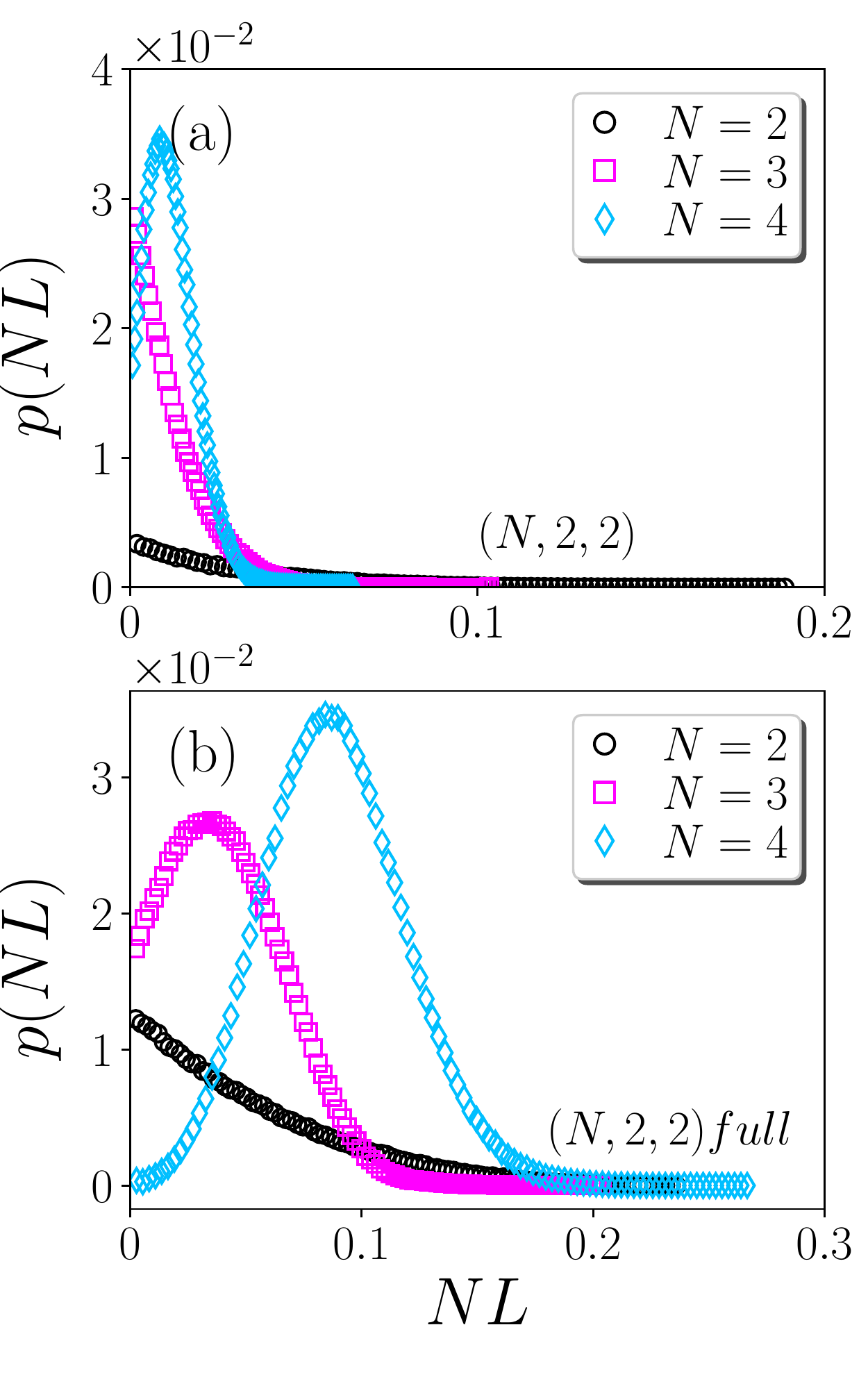}
 \caption{Distance distribution for the nonlocal correlations in the Bell scenario with $10^6$ samples. $a)$  $(N,2,2)$ Bell scenario for $N=2,3,4$. $b)$ $(N,2,2)$ Bell scenario where only full correlators are considered for $N=2,3,4$. The numerical precision is within of $10^{-10}$. We are considering local if the result of $\mathrm{NL}(\q) \leq 10^{-10}$ and nonlocal otherwise.}
 %\label{fig:2m2full}
 \label{fig:N22}
\end{figure}

The distance distribution for the  nonlocal correlations for $N=2,3,4$ is shown in Fig. \ref{fig:N22}a.

%%%
\subsubsection{Full-correlation $(N,2,2)$ Bell scenario }

Also in the $(N,2,2)$ scenario we can  describe the correlations using only full correlations $\left\langle x_1x_2 \ldots x_N \right\rangle$. In this  framework, the nonsignalling  set is the hypercube $\left\{-1,1\right\}^{2^N}$.

To estimate the relative volume of the local set we sample uniformly from the hypercube and calculate the trace distance $\mathrm{NL}$. The results are shown in Table \ref{table:N22full}.

%\begin{figure}
 %\centering
 %\includegraphics[scale=0.4]{N22_nonlocal_histogram.eps}
% \caption{Distance distribution for the  nonlocal behaviours in the full correlation $\left(N,2,2\right)$ Bell scenario for $N=2,3,4,5$. We have sampled $10^{6}$ with numerical precision of $16$ decimal places. We have considered local if $\mathrm{NL}(\q) \leq 10^{-10}$ and nonlocal otherwise. We can see that $N\to \infty$ \sam{Eu coloquei o resultado de $N=5$ para para ilustrar o comportamento que esperamos para $N\to \infty$ (no entanto n\~ao temos estat\'istica suficiente para uma curva suave), $N>4$ demanda um tempo computacional muito grande. Podemos ver que que a curva se afasta muito do ponto zero. Mostrando que \'e nesse limite \'e o volume relativo do local \'e insignificante. }}
 % \label{fig:N22full}
%\end{figure}

%\begin{figure}
 %\centering
 %\includegraphics[scale=0.4]{VLocal_N.eps}
% \caption{How the volume of the local set decreases with $N$ in the $\left(N,2,2\right)$ Bell scenario for $N=2,3,4,5$.}
  %\label{fig:N22full_vol}
%\end{figure}

\begin{table}[h!]
\centering
\begin{tabular}{|l|l|l|}
\hline
$N$& $\sharp L$     & $\%L$          \\ \hline
2 &   666657 & 66.66    \\ \hline
3 &  102367 &  10.23 \\ \hline
4 &  188  &   0.0188 \\ \hline
\end{tabular}
\caption{Relative volume of the classical set in the full-correlation $\left(N,2,2\right)$  Bell scenario using a sample of $10^6$
nonsignalling points.}
\label{table:N22full}
\end{table}

The distance distribution for the nonlocal correlations for $N=2,3,4$ is shown in Figs.
\ref{fig:N22}b. As in the scenario $(2,m,2)$, the volume of the local set decays more 
rapidly when only the full correlators are considered (in comparison with when the 
marginal information is taken into account). And once more, the nonlocal points 
concentrate at a distance that increases as we increase the number of parts $n$.

\subsubsection{$\left(2,2,d\right)$ Bell scenario}

In the $\left(2,2,d\right)$ Bell scenario Alice and Bob have each two $d$-outcome measurements. 
The set of nonsignalling correlations is a $4\left(d^2-d\right)$-dimensional set in $\mathds{R}^{4d^2}$. 
In order to perform the sampling in $\mathcal{C}_{NS}$, we need to find a full-dimensional parametrization of $\mathcal{C}_{NS}$ in $\mathds{R}^{4\left(d^2-d\right)}$.
In this case, we can not use expectation values to describe the correlations and we have carefully chosen which probabilities $p\left(ab \vert xy\right)$ we will keep and which probabilities we will discard.

In this scenario it suffices to keep the probabilities $p\left(ab \vert xy\right)$ with $a=0, \ldots , d-2$ and $b=0, \ldots, d-1$ if $xy=00$ or $xy=11$ and the probabilities $p\left(ab \vert xy\right)$ with $a=0, \ldots , d-1$ and $b=0, \ldots, d-2$ if $xy=01$ or $xy=10$. All other probabilities $p\left(ab\vert xy\right)$ can be recovered from these using nonsignalling and normalization conditions.
This gives a full-dimensional parametrization of $\mathcal{C}_{NS}$ in $\mathds{R}^{4\left(d^2-d\right)}$.
The image of the nonsignalling set under this projection belongs to the hypercube 
$[0,1]^{4\left(d^2-d\right)}$. A point in the hypercube belongs to the image of the 
nonsignalling set if every $p\left(ab\vert xy\right)$ recovered from it is positive.

\begin{figure}
 \centering
 \includegraphics[scale=0.45]{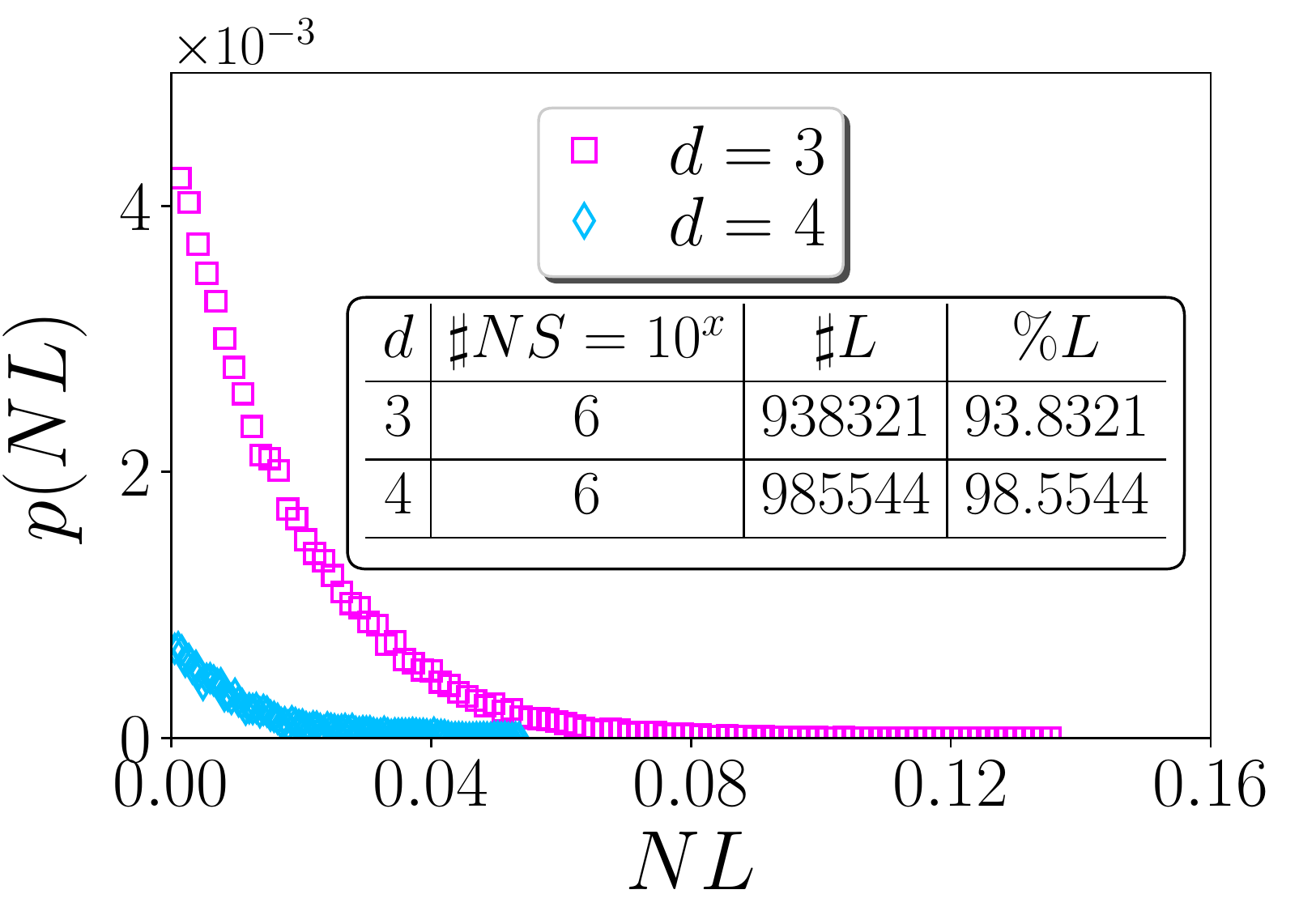}
 \caption{Distance distribution for the nonlocal  correlations in the $(2,2,d)$ Bell scenario for $d=3$ and $4$. }
  \label{fig:22d}
\end{figure}

To estimate the relative volume of the classical set we use the Gibbs
sampler within the \emph{crnpd} Matlab function  and calculate  trace distance $\mathrm{NL}(\q)$. 
The distance distribution for the nonlocal correlations for $d=3,4$ and the relative volumes are shown in Fig.~\ref{fig:22d}. 

%\begin{table}[h!]
%\centering
%\begin{tabular}{|l|l|l|l|l|}
%\hline
%$d$ & $\sharp NS=10^x$ & $\sharp L$     & $\%L$          \\ \hline
%2 & 7 & 9414201 & 94.14201  \\ \hline 
%3 & 6 & 938321 & 93.8321 \\ \hline
%4 & 6 & 985544  &  98.5544  \\ \hline
%
%\end{tabular}
%\caption{Relative volume of the classical set in the $\left(2,2,d\right)$ scenario. \sam{vai deixar o resultado do 222 nessa tabela?}}
%\label{table:22d}
%\end{table}

%%%

Counter-intuitively and contrary to the scenarios $(2,m,2)$ and $(N,2,2)$,
apparently the volume of the local set is increasing and the concentration of the nonlocal points is very close to the local boundary. 
However, these results might be an artifact of the numerical precision. Our codes have a precision of $10^{-10}$ when
computing the trace distance to the local set. That is, any nonlocal point with a distance below this threshold will count as local. 
This is not an issue in the scenarios $(2,m,2)$ and $(N,2,2)$ because not only the local set shrinks but also the distance to it increases. 

Within our numerical precision, two opposite results become possible. In the first, the volume of the local set indeed and
counter-intuitively increases with $d$. The second option is that the local volume decreases but at the same time all the nonlocal points 
concentrate at a very small distance form the local set and thus cannot be distinguished by our numerical precision. In either case, we believe 
that is a very interesting and non-trivial result.

\section{Discussion} \label{sec:Discussion}

Nonlocality is an key concept in foundations of quantum physics and  has also found 
practical  applications. In both contexts, understanding the geometry of the 
nonsignalling, quantum and local set 
is certainly an important primitive. Here we have studied their relative volumes and 
the distribution of a quantitative measure of nonlocality using both analytical and 
numerical methods.

We have found analytically two different classes of full-correlation Bell scenarios 
where the nonsignalling correlations can behave very differently: on the $(2,m,2)$ 
scenario, the correlations are
generically quantum and nonlocal while on the cycle scenario the correlations are generically classical and local. 
Our numerical findings show that a similar result holds true when the entire probability distribution (and not just the full correlators)
are taken into account, and also when considering the scenario $(n,2,2)$ with more 
parts.

%Using numerical methods we have also witnessed the concentration of the volume close to the 
%local set for $(2,2,d)$ scenarios and the  concentration of the volume on the nonlocal portion of the nonsignalling set
%for $(N,2,2)$ scenarios.

The distribution of our chosen quantifier for nonlocality --based on the trace 
distance between the probability
distribution under test and the set of local  correlations introduced in Ref. 
\cite{Brito2018}-- has unveiled interesting features in various scenarios. Of 
particular relevance we have seen that in the scenarios $(2,m,2)$ and $(N,2,2)$ not 
only the volume of the local set decreases very rapidly but also that the nonlocal 
points concentrate at a distance from the local set that increases with both $m$ and 
$N$. We believe that such a surprising behaviour reflects the signature of the 
complicated geometry of the nonsignalling and local sets, giving further insight on 
the relation between local and nonsignalling set.

Regarding the scenario $(2,2,d)$, due to numerical precision, our results are inconclusive so far. 
Considering a precision of $\epsilon=10^{-10}$  in the calculation of the nonlocality quantifier $\mathrm{NL}(\q)$ 
we have seen that the volume of points with $\mathrm{NL}(\q) \leq \epsilon$ increases as we increase the number of outcomes $d$. 
Two options are available. First, the volume of the local set, as opposed to the other scenarios, increases with $d$. This is similar to the 
recent result obtained in \cite{Fonseca2018}, where it is shown that the probability of violation of a Bell inequality in $(2,d,2)$ 
decreases with $d$ if one considers a maximally entangled state and  randomly sampled projective measurements. Second option is that the local volume is indeed decreasing (as intuition would suggest) but at the same time the distribution of $\mathrm{NL}(\q)$ concentrates at an $\epsilon$ distance from the local set. Either way, this shows the complex geometry of the Bell correlations and is a point that certainly deserves further investigation.

Finally, we highlight that even though here we have focused on Bell nonlocality, the  
same 
methods can also be applied not only to study the differences between classical  and 
nonclassical resources in more general notions of nonlocality and  contextuality, but 
also to study the relative volumes --in many different scenarios-- between the 
classical set and those sets of correlations defined by the NPA hierarchy, as 
suggested by the authors in~\footnote{ L. Yeong-Cherng et. al., \textit{Private 
Communication}}. We 
hope our results might motivate further research in these directions.

\begin{acknowledgments}
The authors acknowledge the Brazilian ministries
MEC and MCTIC, funding agency CNPq (PQ grant No.
307172/2017-1 and INCT-IQ.
BA also thanks T. A. Jorge for the help with MATLAB codes. During a stage of elaboration 
of this manuscript CD was also supported by a fellowship from the Grand Challenges 
Initiative at Chapman University.
\end{acknowledgments}

\bibliography{biblio}

\end{document}